\documentclass[osajnl2,showpacs,amsmath,twocolumn]{revtex4}

\makeatletter
\def\Url@Hook{$} %$
\makeatother

\DeclareMathSymbol{\Phi}{\mathalpha}{letters}{8}

\DeclareMathOperator{\e}{e}
\DeclareMathOperator{\RE}{Re}
\DeclareMathOperator{\IM}{Im}
\newcommand{\I}{\mathrm{i}}
\newcommand{\diff}{\mathrm{d}}
\newcommand{\mat}[1]{\hat{#1}}
\newcommand{\vect}[1]{\mathbf{#1}}
\newcommand{\abs}[1]{\lvert #1 \rvert}
\newcommand{\tsub}[1]{_{\text{#1}}}
\newcommand{\tsup}[1]{^{\text{#1}}}
\newcommand{\Vector}[2]{\begin{bmatrix} #1 \\ #2 \end{bmatrix}}
\newcommand{\Matrix}[4]{\begin{bmatrix} #1 & #2 \\ #3 & #4 \end{bmatrix}}

\allowdisplaybreaks[1]

\begin{document}

\title{Application of the multiple-scattering method to analysis of
  systems with semi-infinite photonic waveguides}
\author{Wojciech \'Smigaj} 
\affiliation{Surface Physics Division, Faculty of Physics,
  Adam Mickiewicz University,\\
  Umultowska 85, 61-614 Pozna\'n, Poland}
\email{achu@hoth.amu.edu.pl}

\begin{abstract}
  We propose a technique of compensating the spurious reflections
  implied by the multiple-scattering (MS) method, commonly used for
  analyzing finite photonic crystal (PC) systems, to obtain exact
  values of characteristic parameters, such as reflection and
  transmission coefficients, of PC functional elements.  Rather than a
  modification of the MS computational algorithm, our approach
  involves postprocessing of results obtained by the MS method. We
  derive analytical formulas for the fields excited in a finite
  system, taking explicitly into account the spurious reflections
  occurring at the artificial system boundaries.  The intrinsic
  parameters of the investigated functional element are found by
  fitting the results of MS simulations to those obtained from the
  formulas derived. Devices linked with one and two semi-infinite
  waveguides are analyzed explicitly; possible extensions of the
  formalism to more complex circuits are discussed as well. The
  accuracy of the proposed method is tested in a number of systems;
  the results of our calculations prove to be in good agreement with
  those obtained independently by other authors.
\end{abstract}

\maketitle

\section{Introduction}

Photonic crystals (PCs) have recently become the object of increased
interest as possible hosts for optical functional devices, e.g., beam
splitters, demultiplexers, etc.~\cite{ThylenCPC04} These elements, as
well as the basic building blocks of photonic integrated
circuits---waveguide bends, junctions etc.---have often been
investigated by methods designed for finite systems, such as the
finite-difference time-domain (FDTD) method~\cite{TafloveBook95} and
the multiple-scattering (MS)
method~\cite{FelbacqJOSAA94,TayebJOSAA97}.  Consequently, the devices
in question were considered to be embedded in a finite fragment of a
PC.  This, however, involved spurious reflections at the artificial PC
boundaries, significantly complicating the analysis of the device
behavior.

To remedy this situation, several methods suitable for analyzing
systems with semi-infinite waveguides have been developed.  In the
effective discrete equations method~\cite{MingaleevJOSAB02} and the
Wannier function method~\cite{BuschJPCM03}, the electric and magnetic
fields are represented in a basis of states localized at elementary
defects; various grating-based techniques, like those presented in
\cite{BottenPRE04,WhitePRE04,LiPRB03}, have been developed as well.
Both approaches, however, impose some restrictions on the problems to
which they can be applied: the former is ill-suited to open systems
(i.e., those with vacuum regions extending to infinity), while the
latter only applies to systems with unidirectional waveguides.
Recently, the multiple multipole method has been extended by Moreno
\textit{et al.}~\cite{MorenoPRE02} to systems with semi-infinite
waveguides; here, the fields on a transverse section of each
waveguide, sufficiently distant from discontinuities (junction, bend
etc.), are matched to a linear combination of the waveguide
eigenstates. As demonstrated in~\cite{MorenoPRE02}, this technique is
very general and applicable also to the problems which cannot be dealt
by the other above-mentioned methods.

The aim of this paper is to show how to take into account the presence
of semi-infinite waveguides in calculations performed on the basis of
the MS technique. Owing to its particular simplicity and efficiency in
dealing with the case of PCs composed of cylindrical rods, this
technique has gained significant popularity
\cite{YonekuraJLT99,OchiaiPRB02,ChenPRE04}. Although the approach
proposed by Moreno \textit{et al.}~\cite{MorenoPRE02} could be
straightforwardly carried over to the combination of the MS method and
the method of fictitious sources, formulated in~\cite{TayebJOSAA04},
this would require significant changes in the computational procedure,
as well as availability of externally calculated data describing the
fields corresponding to the waveguide eigenstates. In contrast, the
approach outlined below consists solely in postprocessing of the
results obtained by the `pure' MS method.  While admittedly less
general than the other technique, it is nevertheless applicable to a
number of situations frequently encountered in practice. Two of them
are discussed in Sections \ref{sec:OneWaveguide}
and~\ref{sec:TwoWaveguides}, in which the proposed procedure is
applied to devices linked with one waveguide and two waveguides,
respectively. Possible extensions of the method are discussed in
Section~\ref{sec:Extensions}.

\section{One waveguide}
\label{sec:OneWaveguide}

\subsection{Theory}
\label{subsec:OneWaveguideTheory}

\begin{figure}
  \includegraphics{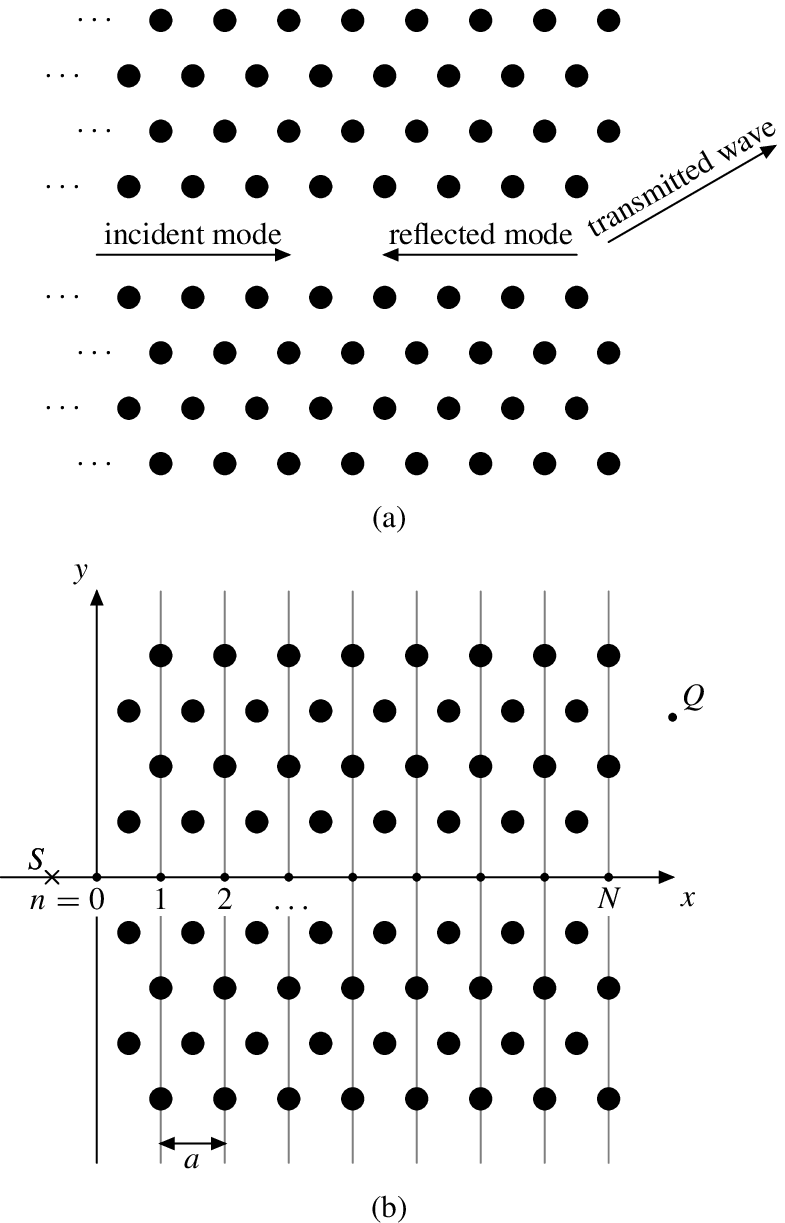}
  \caption{(a) The ideal system: a semi-infinite waveguide terminated
    at the surface of a PC. A single mode propagates along the
    waveguide from inside the PC; on reaching the waveguide outlet, it
    is partially reflected and partially transmitted into free space.
    (b) A finite counterpart of the system shown in (a), amenable to
    analysis by the MS method. The grey lines delimit the $N$~unit
    cells of the waveguide. Smigaj-1.eps\label{fig:System1}}
\end{figure}

The first system to be considered is a single waveguide terminated at
the surface of a semi-infinite PC [Fig.~\ref{fig:System1}(a)]. A
frequently discussed problem (see, e.g.,
\cite{MekisJLT01,HakanssonJLT05,DossouOptComm06}) is the design of the
precise shape of the waveguide outlet that would allow maximization of
the power transmitted into free space. In this case, the main
objective is to calculate the reflection coefficient of the outlet;
the intensity of the field produced at some point of the free-space
region (possibly at infinity) when an eigenmode of unitary power
propagates towards the end of the waveguide is sometimes searched for,
too. In the following we will show how both quantities can be found on
the basis of calculations done for the finite system shown in
Fig.~\ref{fig:System1}(b), presenting a waveguide that is $N$~unit
cells long.

The waveguide is assumed to be a single-mode one, and its unit cell to
have a mirror-symmetry plane parallel to the $yz$ plane, so that the
propagating Bloch states of the waveguide are characterized by
wave numbers $k$ and $-k$. An arbitrary source~$S$ placed near the
waveguide inlet excites the right-propagating mode, which then
undergoes multiple reflections. Let $u_n$ and $d_n$ ($n = 0, 1,
\dotsc, N$) denote the $z$ component of the electric (TM polarization)
or magnetic (TE polarization) field corresponding to the mode of
wave number $k$ and $-k$, respectively, at point~$n$ represented by
coordinates $(x=na, y=y_0)$ in Fig.~\ref{fig:System1}(b). The choice
of $y_0$ is arbitrary; the zero value can be assumed unless the
waveguide eigenmode is antisymmetric with respect to the $xz$ plane.
From the Bloch theorem, we have
\begin{subequations}
  \label{eq:ABloch}
  \begin{align}
    \Vector{u_n}{d_n} &= \mat T^n \Vector{u_0}{d_0}, \\
    \Vector{u_N}{d_N} &= \mat T^{N-n}
    \Vector{u_n}{d_n},
  \end{align}
\end{subequations}
where
\begin{equation}
  \label{eq:ATmatrix}
  \mat T =
  \Matrix{\e^{\I k a}}{0}{0}{\e^{-\I k a}}
  \equiv \Matrix{\Phi}{0}{0}{\Phi^{-1}}
\end{equation}
is the waveguide transfer matrix; $a$~is the waveguide period and
$\Phi \equiv \e^{\I k a}$. The outgoing modes get partly reflected at
the waveguide ends; this can be expressed by the following `boundary
conditions':
\begin{subequations}
  \label{eq:ABndConds}
  \begin{align}
    d_N &= r u_N, \\
    u_0 &= u\tsub{inc} + r' d_0,
  \end{align}
\end{subequations}
where $r$ and $r'$ are the reflection coefficients at the waveguide
outlet and inlet, respectively, and $u\tsub{inc}$ stands for the
effective field corresponding to the right-propagating mode excited by
the source~$S$, extrapolated to point $n=0$. By combining Eqs.\
\eqref{eq:ABloch}--\eqref{eq:ABndConds} and eliminating the variables
$u_0$, $d_0$, $u_N$, and $d_N$, we get the linear system
\begin{equation}
  \label{eq:ALinSystem}
  \Matrix{\Phi^{-n}}{-r' \Phi^n}{r \Phi^{N-n}}{-\Phi^{n-N}}
  \Vector{u_n}{d_n} = \Vector{u\tsub{inc}}{0},
\end{equation}
whose solution reads
\begin{equation}
  \label{eq:ASolution}
  \Vector{u_n}{d_n} =
  \frac{u\tsub{inc}}{1 - r r' \Phi^{2N}}
  \Vector{\Phi^n}{r \Phi^{2N-n}}.
\end{equation}
In cells lying sufficiently far from the waveguide ends for the
contribution of the evanescent states to be negligible, the total
field~$f_n$ can be expressed only in terms of the propagating modes:
\begin{equation}
  \label{eq:ATotalField}
  f_n = u_n + d_n = 
  \frac{\Phi^n + r \Phi^{2N-n}}{1 - r r' \Phi^{2N}} u\tsub{inc}.
\end{equation}
We are interested in the reflection coefficient~$r$, as well as in the
intensity $\abs{f\tsup{ideal}(Q)}^2$ of the field which would be
generated at some point~$Q$ in the free space if the waveguide were
semi-infinite and the incident right-propagating mode carried unitary
power.

The reflection coefficient can be found simply by least-squares
fitting of the numerically calculated values of $f_{n+1}/f_n$ to those
calculated by the formula resulting from Eq.~\eqref{eq:ATotalField}:
\begin{equation}
  \label{eq:AStep1}
  \frac{f_{n+1}}{f_n} = 
  \frac{\Phi^{n+1} + r \Phi^{2N-n-1}}{\Phi^n + r \Phi^{2N-n}}.
\end{equation}
In addition to~$r$, this gives also the value of~$\Phi$, which can be
used for calculating the wave number~$k$. Let us note by the way that
solving Eq.~\eqref{eq:AStep1} for $n = l-1$ and $n=l$, where $l$~is
some fixed integer, yields
\begin{subequations}
  \label{eq:AStartingPt}
  \begin{gather}
    k = \pm\frac{1}{a} \arccos\frac{f_{l-1} + f_{l+1}}{2f_l} 
    \intertext{and}
    r = \frac{f_{l-1}\Phi - f_l}{f_l\Phi -
      f_{l-1}}\Phi^{2(l-N)-1};
  \end{gather}
\end{subequations}
these values, calculated for $l \approx N/2$, i.e., near the waveguide
center, can be used as starting points in the nonlinear least-squares
fitting procedure. The sign of~$k$ should correspond to the physics of
the problem at hand.

It should be stressed that only cells distant enough from the
waveguide ends (i.e., those labeled $n = B, B+1, \dotsc, N-B$, where
the `margin'~$B$ is a sufficiently large integer) should be taken into
account in the above fitting procedure, since Eq.~\eqref{eq:AStep1}
has been derived with the assumption that evanescent states are of
negligible amplitude in the cells labeled $n$ and $n+1$.

Let us proceed to the determination of $\abs{f\tsup{ideal}(Q)}^2$. The
intensity $\abs{f(Q)}^2$ of the field generated at point~$Q$ in the
finite system can be written as
\begin{equation}
  \label{eq:AIntensity}
  \abs{f(Q)}^2 = \abs{t}^2 \abs{\tau(Q)}^2 \abs{u_N}^2 = 
  \frac{\abs{t}^2 \abs{\tau(Q)}^2}{\abs{1 - r r' \Phi^{2N}}^2}
  \abs{u\tsub{inc}}^2,
\end{equation}
where the transmission coefficient $\abs{t}^2 = 1 - \abs{r}^2$
represents the fraction of the total energy emitted into free space
when a waveguide mode reaches the outlet, and the `transfer
coefficient' $\tau(Q)$, dependent on the position of~$Q$ and the
geometry of the waveguide outlet, but not on the waveguide length, is
a measure of the amount of this energy getting to point~$Q$. In the
ideal case of the semi-infinite waveguide, $r'$ would be zero; thus,
\begin{equation}
  \label{eq:AIdealIntensity}
  \abs{f\tsup{ideal}(Q)}^2 = 
  \abs{t}^2 \abs{\tau(Q)}^2 \abs{u\tsub{inc}\tsup{ideal}}^2
\end{equation}
with $\abs{u\tsub{inc}\tsup{ideal}}^2$ chosen so that the incident
mode would carry power $P\tsub{inc}\tsup{ideal} = 1$. From Eqs.\
\eqref{eq:AIntensity}--\eqref{eq:AIdealIntensity} we get
\begin{equation}
  \label{eq:AIdealIntensity2}
  \begin{split}
    \abs{f\tsup{ideal}(Q)}^2 &= \abs{1 - r r' \Phi^{2N}}^2
    \frac{\abs{u\tsub{inc}\tsup{ideal}}^2}{\abs{u\tsub{inc}}^2}
    \abs{f(Q)}^2 \\
    &=\abs{1 - r r' \Phi^{2N}}^2
    \frac{P\tsub{inc}\tsup{ideal}}{P\tsub{inc}} \abs{f(Q)}^2,
  \end{split}
\end{equation}
since the incident power in each case is proportional to
$\abs{u\tsub{inc}}^2$. As shown in the Appendix, the total power
flowing through an arbitrary transverse section of the waveguide is
equal to
\begin{equation}
  \label{eq:APower}
  P = \frac{1 - \abs{r}^2}{\abs{1 - r r' \Phi^{2N}}^2} P\tsub{inc}.
\end{equation}
This power can be easily calculated by the MS method. By using
Eq.~\eqref{eq:APower} for eliminating from
Eq.~\eqref{eq:AIdealIntensity2} the factor $\abs{1 - r r'
  \Phi^{2N}}^2$, which contains the unknown coefficient~$r'$, we
finally get
\begin{equation}
  \label{eq:AIdealIntensityFinal}
  \abs{f\tsup{ideal}(Q)}^2 =
  (1 - \abs{r}^2) \frac{P\tsub{inc}\tsup{ideal}}{P} \abs{f(Q)}^2.
\end{equation}

To sum up, the data obtained in a single MS calculation performed for
a finite system with an $N$-cell-long waveguide suffice for
determination of the reflection coefficient of the waveguide outlet,
as well as the `corrected' field intensity in free space. When only
the former quantity is required, the field excited at $N+1$ sites
lying along the waveguide axis is all that needs to be calculated in
the simulation; otherwise, the power flow through an arbitrary
transverse section of the finite waveguide must be computed too.

\subsection{Examples}

\begin{figure}[b]
  \includegraphics{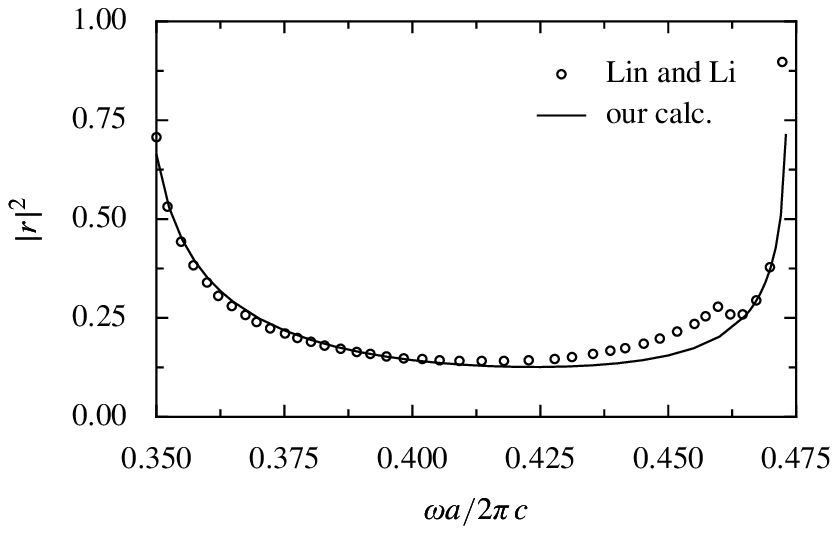}
  \caption{The reflection coefficient of the waveguide outlet shown in
    Fig.~\ref{fig:System1}. Circles: the calculation results of Lin
    and Li~\cite{LinPRB04}; line: results of our
    calculations. Smigaj-2.eps\label{fig:System1ReflCoeff}}
\end{figure}

As an example of application of these results, let us first consider
the system shown in Fig.~\ref{fig:System1}(a): a W1-type waveguide
embedded in a hexagonal lattice of dielectric cylinders with
permittivity $11.56$ and radius $0.18a$, where $a$~is the lattice
constant. The surface termination creates a slight tapering of the
waveguide exit.  Figure~\ref{fig:System1ReflCoeff} presents the
frequency dependence of
the reflection coefficient $\abs{r}^2$ in this configuration; the
results obtained by the proposed technique are compared to the data
reported in~\cite{LinPRB04} [Fig.~7(c)], acquired by the
plane-wave-based transfer-matrix method. Our calculations were done
for waveguide length $N = 15$ with margin $B = 4$.  The convergence is
very fast: in fact, the data obtained in the $N = 7$, $B = 2$ case
(essentially the shortest waveguide for which fitting makes sense)
differ by at most 6\% from those plotted in the graph, and for
frequency values below $0.468 \times 2\pi c/a$ the difference does not
exceed 1\%. There is a good agreement between our results and those of
Lin and Li~\cite{LinPRB04}, except for a small peak at frequency value
$\omega \approx 0.46 \times 2\pi c/a$, present in the
$\abs{r(\omega)}^2$ plot obtained by these Authors, but not reproduced
by the curve resulting from our calculations. Since the waveguide is a
single-mode one and the mode dispersion curve is smooth around this
frequency value, the physical origin of this sharp peak is not clear
to us and we believe it might be a numerical artifact.

\begin{figure}
  \includegraphics{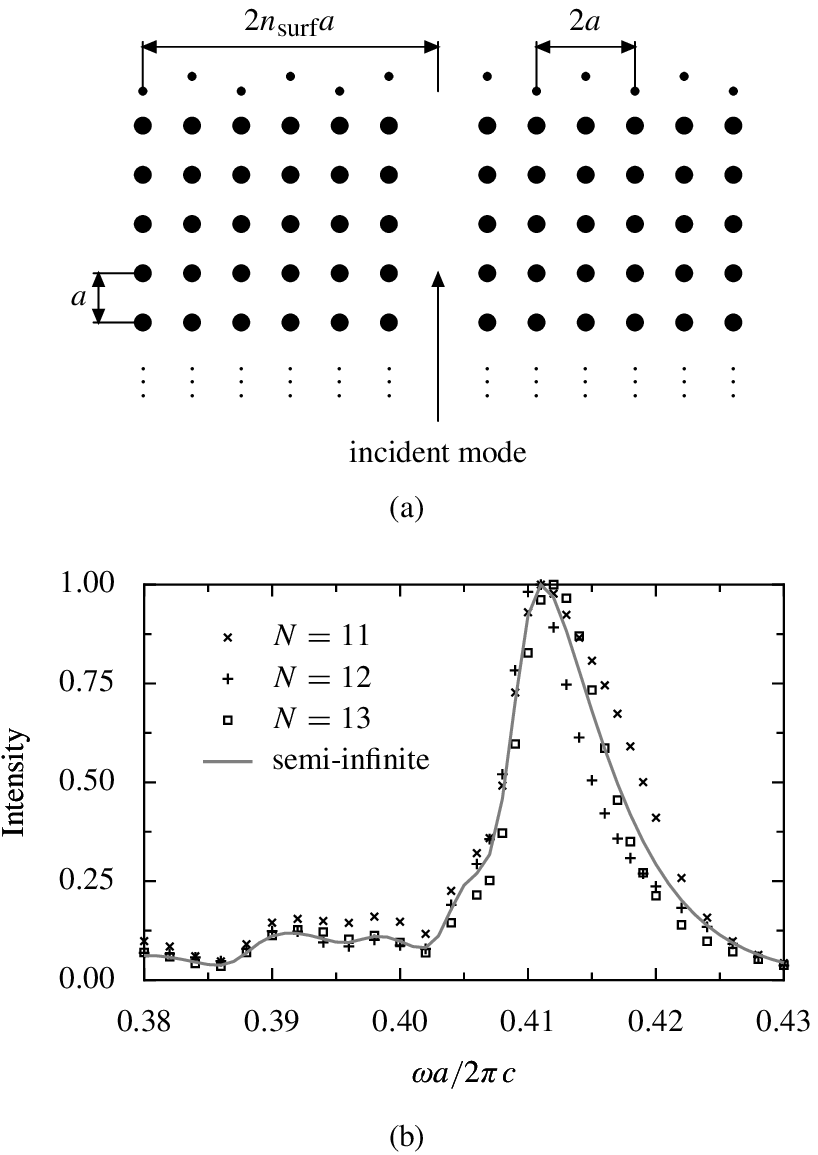}
  \caption{(a) A photonic crystal leaky-wave antenna. The radii of the
    bulk and the surface cylinders are $0.18a$ and $0.09a$,
    respectively, $a$~being the lattice constant; all cylinders have
    permittivity $11.56$. Every second surface cylinder is shifted by
    $0.3a$ towards the bulk. The corrugated surface is $n\tsub{surf} =
    9$ unit cells long. (b) The far-field intensity of the radiation
    emitted perpendicularly to the surface, calculated for waveguide
    length $N = 11, 12$, and $13$, as well as for the semi-infinite
    waveguide on the basis of data obtained for $N = 12$, $B = 3$.
    Normalization: see text. Smigaj-3.eps\label{fig:Antenna}}
\end{figure}

As a second example, let us consider the leaky-wave photonic
antenna \cite{MorenoPRB04,KramperPRL04,Smigaj06} shown in
Fig.~\ref{fig:Antenna}(a). Here, the
surface surrounding the waveguide outlet is corrugated with period
$2a$ and supports leaky surface modes, which are excited by the
radiation coming from the waveguide. At a certain frequency value
waves scattered at individual perturbed surface cylinders interfere
constructively along the surface normal to produce a collimated beam.
Figure~\ref{fig:Antenna}(b) presents the frequency dependence of the
far-field intensity of the radiation emitted perpendicularly to the
surface; the outcomes of `na\"\i{}ve' calculations done for
finite-length waveguides are juxtaposed with those obtained by the
proposed scheme for the semi-infinite waveguide.  All curves are
normalized to their absolute maxima. Evidently, in the
finite-waveguide case, the shape of the main peak depends on the
precise length of the waveguide; this dependence is especially
pronounced to the right of the maximum, where the intensity values
obtained for waveguides 11 and~12 cells long differ by as much as a
factor of two. On the other hand, the sidelobes in the frequency range
$\omega < 0.40 \times 2\pi c/a$ are not significantly affected by changes
in the waveguide length; this leads to the conclusion that they result
from interference occurring at the crystal surface rather than in the
waveguide.

\section{Two waveguides}
\label{sec:TwoWaveguides}

\subsection{Theory}

\begin{figure}
  \includegraphics{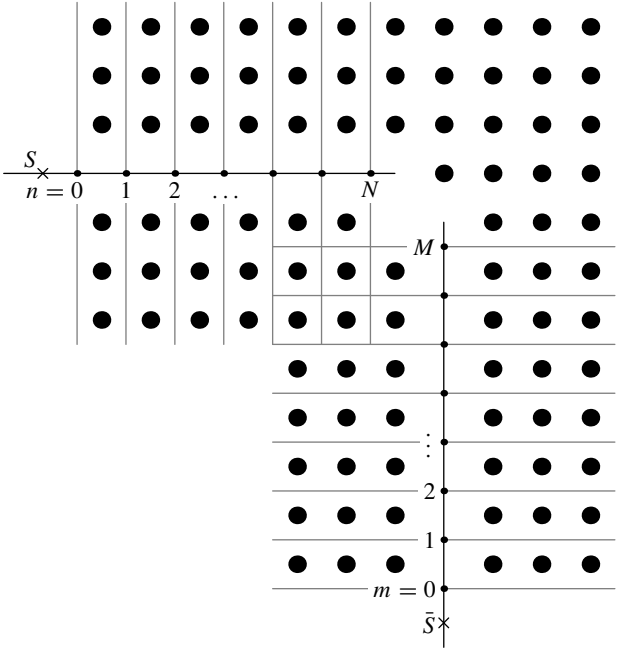}
  \caption{An example system with two waveguides linked by a junction
    (in this case, a bend). Grey lines delimit the unit cells of both
    waveguides. Smigaj-4.eps\label{fig:System2}}
\end{figure}

In this Section, we will focus on the more complex system with two
linked semi-infinite waveguides; the link can be realized, for
instance, by a junction, a bend or a resonant cavity where scattering
can occur. Our goal is to calculate the reflection and transmission
coefficients of this discontinuity. The waveguides are again assumed
to be single-mode ones and to possess a transverse symmetry plane; to
these assumptions let us add that of their identical geometry.
Figure~\ref{fig:System2} depicts the finite system used in numerical
calculations, where the left and right waveguide comprise $N$ and~$M$
unit cells, respectively, and the field is excited by the source~$S$.

Let $u_n$ and $d_n$ denote the field (electric or magnetic, depending
on the polarization), at point $n = 0, 1, \dotsc, N$ inside the left
waveguide in Fig.~\ref{fig:System2}, corresponding to the incoming and
outgoing guided modes, respectively. Their right-waveguide
counterparts at point $m = 0, 1, \dotsc, M$ will be labeled $\bar
u_m$ and $\bar d_m$.  The junction linking the waveguides can be
described by its scattering matrix $\mat S$ defined by
\begin{equation}
  \label{eq:BSJunMatrix}
  \Vector{d_N}{\bar d_M} = 
  \mat S \Vector{u_N}{\bar u_M},
  \qquad
  \mat S \equiv \Matrix{\rho}{\bar\tau}{\tau}{\bar\rho};
\end{equation}
$\rho$, $\bar\rho$ and $\tau$, $\bar\tau$ denote the respective
reflection and transmission coefficients of the junction.  The
`boundary conditions' are in this case
\begin{subequations}
  \label{eq:BBndConds}
  \begin{align}
    \bar u_0 &= r \bar d_0, \\
    u_0 &= u\tsub{inc} + r d_0,
  \end{align}
\end{subequations}
where $r$ is the reflection coefficient at the waveguide outer ends
(the waveguides having the same geometry, their terminations can be
assumed to be identical too), and $u\tsub{inc}$ is defined as in
Section~\ref{sec:OneWaveguide}.  The Bloch theorem gives
\begin{align}
  \label{eq:BBloch1}
  \Vector{u_n}{d_n} &= \mat T^n \Vector{u_0}{d_0}, &
  \Vector{\bar u_m}{\bar d_m} &= \mat T^m \Vector{\bar u_0}{\bar d_0}, \\
  \label{eq:BBloch2}
  \Vector{u_N}{d_N} &= \mat T^{N-n} \Vector{u_n}{d_n}, &
  \Vector{\bar u_M}{\bar d_M} &= \mat T^{M-m} \Vector{\bar u_m}{\bar d_m},
\end{align}
with the waveguide transfer matrix $\mat T$ defined, as before, by
Eq.~\eqref{eq:ATmatrix}. By eliminating the variables $u_0$, $d_0$,
$\bar u_0$ and $\bar d_0$ from Eqs.\ \eqref{eq:BBndConds}
and~\eqref{eq:BBloch1}, and the variables $u_N$, $d_N$, $\bar u_M$,
and $\bar d_M$ from Eqs.\ \eqref{eq:BSJunMatrix}
and~\eqref{eq:BBloch2}, we obtain the linear system
\begin{widetext}
  \begin{equation}
    \label{eq:BMainLinSystem}
    \begin{bmatrix}
      \Phi^{-n} & -r \Phi^n & 0 & 0 \\
      0 & 0 & \Phi^{-m} & -r \Phi^m \\
      \rho \Phi^{N-n} & -\Phi^{-N+n} & \bar\tau\Phi^{M-m} & 0 \\
      \tau \Phi^{N-n} & 0 & \bar\rho \Phi^{M-m} & - \Phi^{-M+m} 
    \end{bmatrix}
    \begin{bmatrix}
      u_n \\ d_n \\ \bar u_m \\ \bar d_m
    \end{bmatrix}
    = 
    \begin{bmatrix}
      u\tsub{inc} \\ 0 \\ 0 \\ 0
    \end{bmatrix} 
    ,
  \end{equation}
  the solution of which yields the expressions for the total fields in
  the left and the right waveguides:
  \begin{subequations}
    \label{eq:BfSolution}
    \begin{align}
      f_n &\equiv u_n + d_n 
      = \frac{(1 - r \bar\rho \Phi^{2M}) \Phi^n + [\rho + r(\tau\bar\tau -
        \rho\bar\rho) \Phi^{2M}] \Phi^{2N-n}}
      {(1 - r \bar\rho \Phi^{2M}) - r
        [\rho + r(\tau\bar\tau - \rho\bar\rho) \Phi^{2M}] \Phi^{2N}}
      u\tsub{inc}, \\
      \bar f_m &\equiv \bar u_m + \bar d_m 
      = \frac{\tau(r\Phi^m + \Phi^{-m})\Phi^{N+M}} 
      {(1 - r \bar\rho \Phi^{2M}) - r
        [\rho + r(\tau\bar\tau - \rho\bar\rho) \Phi^{2M}] \Phi^{2N}}
      u\tsub{inc}.
    \end{align}
  \end{subequations}
\end{widetext}

The above procedure can be repeated for the situation when source
$S$~is replaced with source~$\bar S$ located near the entry of the
\emph{right} waveguide; in this case, the fields $g_n$ and~$\bar g_m$
in the left and the right waveguides are given by
\begin{subequations}
  \label{eq:BgSolution}
  \begin{align}
    g_n &=
    \frac{\bar\tau(r\Phi^n + \Phi^{-n})\Phi^{N+M}} 
    {(1 - r \rho \Phi^{2N}) - r
      [\bar\rho + r(\tau\bar\tau - \rho\bar\rho) \Phi^{2N}] \Phi^{2M}}
    \bar u\tsub{inc}. \\
    \bar g_m &=
    \frac{(1 - r \rho \Phi^{2N}) \Phi^m + [\bar\rho + r(\tau\bar\tau -
      \rho\bar\rho) \Phi^{2N}] \Phi^{2M-m}}
    {(1 - r \rho \Phi^{2N}) - r
      [\bar\rho + r(\tau\bar\tau - \rho\bar\rho) \Phi^{2N}] \Phi^{2M}}
    \bar u\tsub{inc},
  \end{align}
\end{subequations}
$\bar u\tsub{inc}$ being defined analogously to $u\tsub{inc}$.

It is important to note that the MS method allows very efficient
calculation of the fields excited in a single structure by several
independent sources (e.g., $S$ and~$\bar S$), since the scattering
matrix of the whole system, whose diagonalization is by far the most
time-consuming step of the computational algorithm, is independent of
the incident field~\cite{FelbacqJOSAA94,TayebJOSAA97}. In the
following we show how the values of $f_n$, $\bar f_m$, $g_n$, and
$\bar g_m$, calculated by the MS method, can be used for determination
of the junction parameters $\rho$, $\bar\rho$, $\tau$, and $\bar\tau$.
As before, in all these computations only cells sufficiently distant
from the waveguide ends should be taken into account.

\begin{enumerate}
\item We have
  \begin{equation}
    \label{eq:BStep1}
    \frac{\bar f_{m+1}}{\bar f_m} =
    \frac{r\Phi^{m+1} + \Phi^{-m-1}}{r \Phi^m + \Phi^{-m}};
  \end{equation}
  therefore, as in the one-waveguide case, the parameters $\Phi$
  and~$r$ can be obtained by least-squares fitting of the above
  formula's right-hand side to simulation results. Good starting
  points for the fitting procedure are in this case
  \begin{subequations}
    \label{eq:BStartingPt}
    \begin{gather}
      k = \pm\frac{1}{a}\arccos\frac{\bar f_{l-1} + \bar f_{l+1}}{2\bar f_l} 
      \intertext{and}
      r = \frac{\bar f_l\Phi - \bar f_{l-1}}{\bar f_{l-1}\Phi - \bar f_l}
      \Phi^{-2l-1}
    \end{gather}
  \end{subequations}
  with $l \approx M/2$.

\item To shorten the notation, we introduce the following symbols:
  \begin{align}
    \label{eq:BxDef}
    \mu &\equiv r \rho \Phi^{2N}, &
    \bar \mu &\equiv r \bar\rho \Phi^{2M},\\
    \label{eq:ByDef}
    \nu &\equiv r \tau \Phi^{2N}, &
    \bar \nu &\equiv r \bar\tau \Phi^{2M},\\
    \label{eq:BzwDef}
    \zeta &\equiv \mu \bar \mu - \nu \bar \nu, &
    \eta &\equiv 1 - \mu - \bar \mu + \zeta;
  \end{align}
  consequently, the formulas for $f_n$, $\bar f_m$, $g_n$, and $\bar
  g_m$ become:
  \begin{subequations}
    \label{eq:BfShort}
    \begin{align}
      f_n &= 
      \frac{(1-\bar \mu) \Phi^n + r^{-1}(\mu-\zeta)\Phi^{-n}}{\eta} 
      u\tsub{inc}, \\
      \bar f_m &= 
      \frac{\nu \Phi^m + r^{-1}\nu\Phi^{-m}}{\eta} \Phi^{M-N} u\tsub{inc}, \\
      g_n &= 
      \frac{\bar \nu \Phi^n + r^{-1}\bar \nu\Phi^{-n}}{\eta}
      \Phi^{N-M} \bar u\tsub{inc}, \\
      \bar g_m &= 
      \frac{(1-\mu) \Phi^m + r^{-1}(\bar \mu-\zeta)\Phi^{-m}}{\eta} \bar
      u\tsub{inc}.
    \end{align}
  \end{subequations}
  Thus, \emph{linear} least-squares fitting of the numerically
  calculated values of $f_n$ to the function $\alpha_1 \Phi^n +
  \beta_1 \Phi^{-n}$ allows to find the coefficients
  \begin{subequations}
    \label{eq:BStep2Params}
    \begin{align}
      \alpha_1 &= \frac{1-\bar \mu}{\eta} u\tsub{inc},&
      \beta_1 &= \frac{\mu-\zeta}{\eta} \frac{u\tsub{inc}}{r}.\\
      \intertext{Similarly, fitting the values of
        $\bar f_m$, $g_n$, and $\bar g_m$ to the functions $(\alpha_2
        \Phi^m + \beta_2 \Phi^{-m})\Phi^{M-N}$, $(\alpha_3 \Phi^n +
        \beta_3 \Phi^{-n})\Phi^{N-M}$, and $\alpha_4 \Phi^m + \beta_4
        \Phi^{-m}$, respectively, yields the values of the
        coefficients}
      \alpha_2 &= \frac{\nu}{\eta} u\tsub{inc},&
      \beta_2 &= \frac{\nu}{\eta} \frac{u\tsub{inc}}{r},\\
      \alpha_3 &= \frac{\bar \nu}{\eta} \bar u\tsub{inc},&
      \beta_3 &= \frac{\bar \nu}{\eta} \frac{\bar u\tsub{inc}}{r},\\
      \alpha_4 &= \frac{1-\mu}{\eta} \bar u\tsub{inc},& \beta_4 &=
      \frac{\bar \mu-\zeta}{\eta} \frac{\bar u\tsub{inc}}{r}.
    \end{align}
  \end{subequations}

\item With definitions~\eqref{eq:BzwDef} of $\zeta$ and~$\eta$
  included, the formulas for $\alpha_1$, $\beta_1$, $\alpha_2$,
  $\alpha_3$, $\alpha_4$, and $\beta_4$ form a system of six equations
  with six unknowns: $\mu$, $\bar \mu$, $\nu$, $\bar \nu$,
  $u\tsub{inc}$, and $\bar u\tsub{inc}$. Its solution reads
  \begin{subequations}
    \label{eq:BStep3Solution}
    \begin{align}
      \mu &= \frac{r\alpha_4\beta_1 - \alpha_2\alpha_3}
      {\alpha_1\alpha_4 - \alpha_2\alpha_3},&
      \bar \mu &= \frac{r\alpha_1\beta_4 - \alpha_2\alpha_3}
      {\alpha_1\alpha_4 - \alpha_2\alpha_3},\\
      \nu &= \frac{\alpha_2(\alpha_4 - r\beta_4)}
      {\alpha_1\alpha_4 - \alpha_2\alpha_3},&
      \bar \nu &= \frac{\alpha_3(\alpha_1 - r\beta_1)}
      {\alpha_1\alpha_4 - \alpha_2\alpha_3},\\
      u\tsub{inc} &= \alpha_1 - r\beta_1,&
      \bar u\tsub{inc} &= \alpha_4 - r\beta_4.
    \end{align}
  \end{subequations}

\item The reflection and transmission coefficients $\rho$, $\bar\rho$,
  $\tau$, and $\bar\tau$ can now be calculated from Eqs.\
  \eqref{eq:BxDef}--\eqref{eq:ByDef}, since the values of all the
  other parameters are already known.
\end{enumerate}

\subsection{Example}

\begin{figure}[b]
  \includegraphics{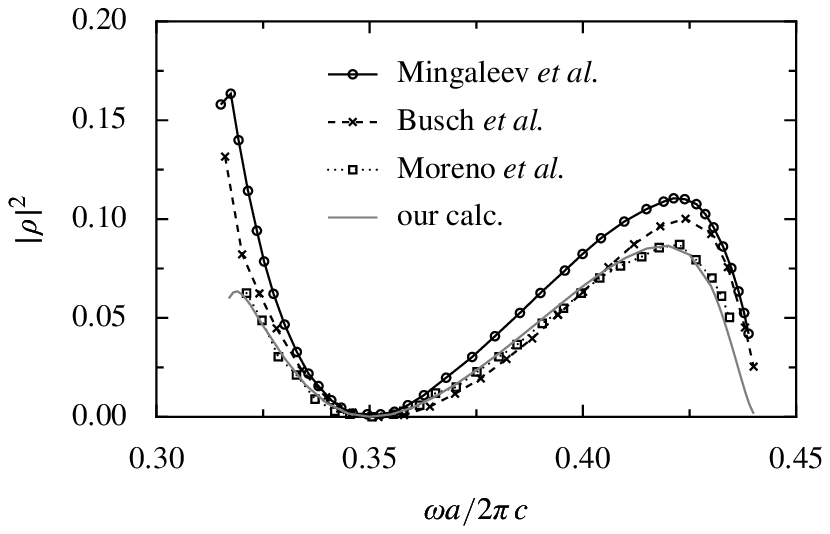}
  \caption{The reflection coefficient of the bend shown in
    Fig.~\ref{fig:System2} calculated by different methods (see text).
    The data plotted with circles, crosses, and squares have been
    taken from \cite{MingaleevJOSAB02}, \cite{BuschJPCM03}
    and~\cite{MorenoPRE02}, respectively.
    Smigaj-5.eps\label{fig:Bend}}
\end{figure}

To test the accuracy of our method, we are going to apply it to the
waveguide bend depicted in Fig.~\ref{fig:System2}. A number of
numerical studies of this system, using different methods, are
available in the
literature \cite{MingaleevJOSAB02,MorenoPRE02,BuschJPCM03}, providing a
reference for new techniques. The superimposed plot in
Fig.~\ref{fig:Bend} shows the frequency dependence of the reflection
coefficient $\abs{\rho(\omega)}^2$ (since the bend is symmetric, $\rho
= \bar\rho$) calculated according to the above-presented scheme
against the spectra obtained by the effective discrete equations
method~\cite{MingaleevJOSAB02}, the Wannier function
method~\cite{BuschJPCM03}, and the multiple multipole
method~\cite{MorenoPRE02}. Clearly, the overall shape of all four
curves is very similar, although the exact values of
$\abs{\rho(\omega)}^2$ differ, especially near the boundaries of the
bulk crystal gap. The spectrum resulting from our calculations is
almost identical to that calculated by Moreno \textit{et
  al}.~\cite{MorenoPRE02} This may be partly due to the similarity of
the MS and multiple multipole methods (in both of them the field is
expanded in a basis of Bessel and Hankel functions); however, it must
also be noted that these methods are intrinsically `exact' in the
sense that the Maxwell equations are solved rigorously, which makes
the calculations accuracy only depend on the maximum order of the
basis functions kept, and the results are known to be rapidly
convergent with the truncation order~\cite{TayebJOSAA97}. In contrast,
the derivation of the effective discrete
equations~\cite{MingaleevJOSAB02} involves a number of approximations
(for instance, only the monopole eigenmodes of the elementary defects
are taken into account). This may be the reason why the reflection
coefficient calculated by Mingaleev \textit{et al.}\ takes values
$\sim30\%$ larger than those obtained in our approach.

\section{Extensions}
\label{sec:Extensions}

The two device classes discussed above, i.e., those linked with one or
two semi-infinite waveguides, represent a large fraction of commonly
investigated photonic building blocks. Though the two-waveguide case
has been analyzed with the assumption of identical geometry of both
waveguides, this condition can be eliminated, with the transmission
coefficients renormalized to refer to eigenmodes carrying unitary
power, in a way similar to that presented for a single waveguide at
the end of Section~\ref{subsec:OneWaveguideTheory}.

Our formalism can be extended to systems with more than two waveguides
as well. However, many such systems, including most \emph{T}
and~\emph{Y} junctions discussed in the literature, are symmetric with
respect to the axis of one or more constituent waveguides. Therefore,
they can also be analyzed with the simpler two-waveguide formalism
presented in this paper. To see this, note that the scattering
matrix~$\mat S$ of a three-waveguide junction is defined by [cf.\
Eq.~\eqref{eq:BSJunMatrix}]
\begin{equation}
  \label{eq:EScattMat}
  \begin{bmatrix}
    d_N^1 \\ d_N^2 \\ d_N^3
  \end{bmatrix}
  =
  \mat S
  \begin{bmatrix}
    u_N^1 \\ u_N^2 \\ u_N^3
  \end{bmatrix}
  \quad\text{with}\quad
  \mat S \equiv
  \begin{bmatrix}
    \rho_1 & \tau_{21} & \tau_{31} \\
    \tau_{12} & \rho_2 & \tau_{32} \\
    \tau_{13} & \tau_{23} & \rho_3
  \end{bmatrix}
  ,
\end{equation}
where $u_N^i$ and $d_N^i$ refer to the incoming and outgoing mode,
respectively, in the $i$th waveguide ($i = 1, 2, 3$). For simplicity
reasons, all the waveguides are assumed to be of the same length~$N$.
If waveguides 2 and~3 are identical and the axis of waveguide~1 is a
symmetry plane of the system, we have $\rho_2 = \rho_3$, $\tau_{12} =
\tau_{13}$, $\tau_{21} = \tau_{31}$, and $\tau_{23} = \tau_{32}$.  To
these equalities we can add $u_N^2 = u_N^3$ and $d_N^2 = d_N^3$,
provided that the sources are placed symmetrically with respect to
that axis. As a result, Eq.~\eqref{eq:EScattMat} simplifies to
\begin{equation}
  \label{eq:EScattMatSimplified}
  \Vector{d_N^1}{d_N^2} = 
  \Matrix{\rho_1}{2\tau_{21}}{\tau_{12}}{\rho_2 + \tau_{23}}
  \Vector{u_N^1}{u_N^2}.
\end{equation}
Therefore, the junction can be treated as a two-waveguide system with
effective reflection and transmission coefficients given by the above
formula. The values of these parameters can be calculated by the
method presented in Section~\ref{sec:TwoWaveguides}. Evidently, the
coefficients $\rho_2$ and $\tau_{23}$ occur only in the form of their
sum, and therefore cannot be obtained independently. However,
normally, the quantities of interest are the coefficients $\rho_1$ and
$\tau_{12}$, which are related to transmission and reflection of the
radiation incoming from the first (`input') waveguide.
Equation~\eqref{eq:EScattMatSimplified} clearly shows that these
parameters can be obtained straightforwardly from the two-waveguide
formalism.

\section{Conclusions}

We have presented a method, based on multiple-scattering numerical
simulations performed for finite systems, allowing to find the
reflection and transmission coefficients of photonic crystal
functional elements linked with ideal semi-infinite waveguides. As our
approach does not involve modification of the existing code
implementing the multiple-scattering method, no serious programming
and testing effort are necessary for its application. The proposed
formalism allows for dealing with a wide variety of photonic crystal
building blocks likely to be encountered in practice, as demonstrated
by its successful application to a tapered waveguide outlet, a
photonic crystal leaky-wave antenna, and a waveguide bend.

\appendix*
\section{Power flow in finite waveguides}

This Appendix presents the derivation of formula~\eqref{eq:APower} for
the power flowing through an arbitrary transverse section of the
waveguide shown in Fig.~\ref{fig:System1}(b). Only TM polarization
will be considered; the procedure to be followed in the case of TE
polarization is completely analogous.

The power~$P(na)$ flowing to the right through a plane $x = na$
(perpendicular to the waveguide axis) per unit length in the
$z$~direction is given by the integral of the $x$~component of the
time-averaged Poynting vector~$\vect{S}$:
\begin{equation}
  \label{eq:CPowerDef}
  \begin{split}
    P(na) &= \int_{-\infty}^\infty S_x \,\diff y \\
    &= \frac{1}{2} 
    \RE \biggl[ \int_{-\infty}^\infty E_z(na, y) \, H_y^*(na, y) \,\diff y
    \biggr].
  \end{split}
\end{equation}
From Eq.~\eqref{eq:ATotalField} we have
\begin{equation}
  \label{eq:CEz}
  E_z(na, y) = \frac{\Phi^n + r \Phi^{2N-n}}{1 - r r' \Phi^{2N}} e_z(y),
\end{equation}
where $e_z(y)$ denotes the electric field corresponding to the
right-propagating waveguide eigenmode of amplitude satisfying the
condition $e_z(0) = u\tsub{inc}$. Similarly, for the magnetic field we
obtain
\begin{equation}
  \label{eq:CHy}
  H_y(na, y) = \frac{\Phi^n - r \Phi^{2N-n}}{1 - r r' \Phi^{2N}} h_y(y);
\end{equation}
note the change of sign in the numerator, which results from the
magnetic field being a pseudovector. By including these formulas into
Eq.~\eqref{eq:CPowerDef}, after some straightforward algebra, we
arrive at
\begin{equation}
  \label{eq:CPowerIntermediate}
  \begin{split}
    P(na) &= \frac{1}{\abs{1 - r r' \Phi^{2N}}^2} \\
    &\quad\times \bigl\{ (1 - \abs{r}^2) \RE J - 2 \IM [r \Phi^{2(N-n)}]
    \IM J \bigr\},
  \end{split}
\end{equation}
where
\begin{equation}
  \label{eq:CJDef}
  J \equiv \frac{1}{2} 
  \int_{-\infty}^\infty e_z(y) \, h_y^*(y) \,\diff y.
\end{equation}
From the conservation of energy, $P(na)$ must be constant for all~$n$,
since the waveguide is not leaky. We conclude that the second term in
braces in Eq.~\eqref{eq:CPowerIntermediate}, being $n$-dependent, must be
identically zero; this leads to the following condition for the fields
corresponding to the waveguide eigenmode:
\begin{equation}
  \label{eq:CCondition}
  \IM J = 0.
\end{equation}
In consequence, Eq.~\eqref{eq:CPowerIntermediate} simplifies to 
\begin{equation}
  \label{eq:PowerFinal}
  P = \frac{1 - \abs{r}^2}{\abs{1 - r r' \Phi^{2N}}^2} P\tsub{inc},
\end{equation}
where $P\tsub{inc} \equiv \RE J$ denotes the power carried by the
incident mode.

% \bibliography{pc}

\end{document}